\documentclass[preprint,aps,pre,showpacs,preprintnumbers,amsmath,amssymb]
{revtex4}

\usepackage{graphicx}

\begin{document}

\title{Pattern formation in crystal growth under parabolic shear flow}

\author{K. Ueno}

\email{ueno@riam.kyushu-u.ac.jp}

\affiliation{Institute of Low Temperature Science, Hokkaido University, Sapporo, 060-0819, Japan}

\altaffiliation[Present address: ]{Research Institute for Applied Mechanics, Kyushu University, 6-1 Kasuga-koen, Kasuga, Fukuoka 816-8580, Japan}


\begin{abstract}
Morphological instability of the solid-liquid interface occuring in a crystal growing from an undercooled thin liquid being bounded on one side by a free surface and flowing down inclined plane is investigated by a linear stability analysis under shear flow. It is found that restoring forces due to gravity and surface tension is important factor for stabilization of the solid-liquid interface on long length scales. This is a new stabilizing effect different from the Gibbs-Thomson effect. A particular long wavelength mode of about 1 cm of wavy pattern observed on the surface of icicles covered with thin layer of flowing water is obtained from the dispersion relation including the effect of flow and restoring forces.
\end{abstract}


\maketitle

\section{introduction}
The interaction between an imposed shear flow and a phase transition underlies a broad range of phenomena \cite{1}. The stability of order under the influence of shear flow is fundamental for engineering to understanding frictional wear \cite{2} and lubrication \cite{3}. In pattern formation in nature, ripple formation in sand induced by water shear flow are well known \cite{4}. In theoretical works, the effect of shear flow on the morphological stability has been studied \cite{5,6,7}. 

An example of morphological instability of the solid-liquid interface in the long wavelength region of about 1 cm under shear flow bounded on one side by a free surface is wavy pattern occuring on the surface of icicles (see Fig. 1 in Ref. \cite{20} and Fig. 9A in Ref. \cite{21}). In its relevant experiment of a crystal growth from a thin liquid flowing down an inclined plane with angle $\theta$ sketched in Fig. \ref{fig:diagram}, it is found that mean wavelength of the wavy pattern of the solid-liquid interface is about $0.83/(\sin\theta)^{0.6\sim0.9}$ cm \cite{22}. Ogawa and Furukawa have recently proposed a model of the mechanism of occurrence of the wavy pattern and obtained reasonable values of wavelength on the icicles \cite{20}. However, in order to explain more quantitatively the experimental result mentioned above, we modify their dispersion relations in the form that includes the effect of restoring forces due to gravity and surface tension on stability of the solid-liquid interface. Furthermore, we improve their formulations by using a linear stability analysis under forced flow developed firstly by Delves \cite{5}. From the dispersion relation in the long wavelength approximation, we present a new amplification rate and phase velocity for the fluctuation of the solid-liquid interface different from that of Ogawa and Furukawa's model.

This paper is organized as follows. In Sec. II, we develop generally the dispersion relation for the fluctuation of the solid-liquid interface. 
In Sec. III, a perturbed normal flow induced by deformation of the solid-liquid interface is determined in the long wavelength approximation. In Sec. IV, a general solution of the perturbed temperature distribution in the liquid is obtained. In Sec. V, we determine the dispersion relation for the fluctuation of the solid-liquid interface in a crystal growth from a thin liquid flowing down an inclined plane by applying the solutions in Sec. III and IV to the general formulation in Sec. II. Section. VI is devoted to discussion. Conclusion is given in Sec. VII. 

\section{dispersion relation for the fluctuation of the solid-liquid interface}
\begin{figure}
\begin{center}
\includegraphics[width=8cm,height=8cm,keepaspectratio,clip]{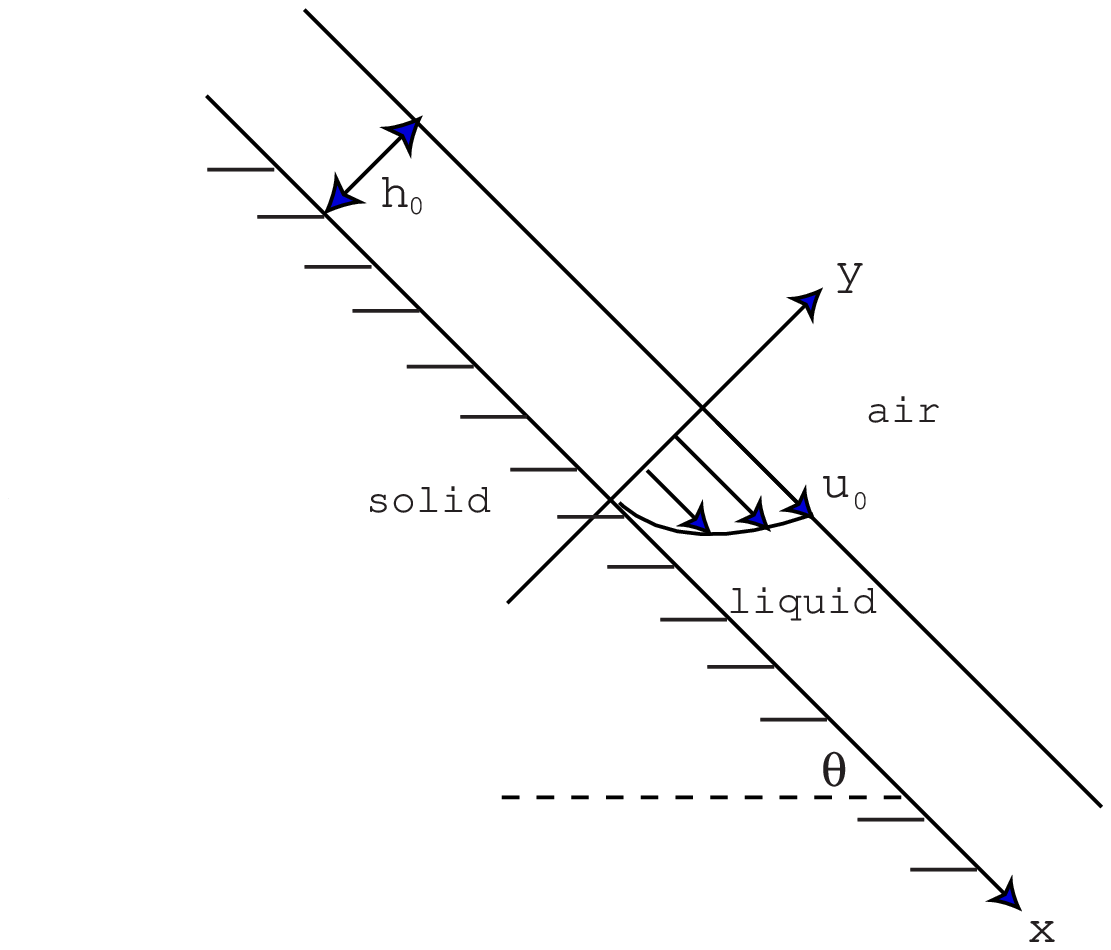}
\end{center}
\caption{Schematic diagram of crystal growth from liquid flowing down inclined plane.}
\label{fig:diagram}
\end{figure}

We consider a crystal growth from an undercooled thin liquid flowing down an inclined plane with the angle $\theta$ \cite{22}. Hereafter the analysis will be restricted to two dimension in a vertical plane $(x,y)$ sketched in Fig. \ref{fig:diagram}. The primary shear flow is parallel to the $x$ axis, and the mean velocity varies only in the $y$ direction. The shear flow is bounded on one side by a free surface. We note that $h_{0}$ is the mean thickness of the liquid, and $u_{0}$ is the velocity at the free surface. In this section, we develop generally the dispersion relation for the fluctuation of the solid-liquid interface by following the ideas \cite{5,25}. 

In the frame of reference moving at the solid-liquid interface velocity $\bar{V}$, the equations for the temperature in the flowing liquid $T_{l}$ and that in the solid $T_{s}$ are
\begin{equation}
\frac{\partial T_{l}}{\partial t}-\bar{V}\frac{\partial T_{l}}{\partial y}+u\frac{\partial T_{l}}{\partial x}
+v\frac{\partial T_{l}}{\partial y}
=\kappa_{l}\left(\frac{\partial^{2} T_{l}}{\partial x^{2}}+\frac{\partial^{2} T_{l}}{\partial y^{2}}\right),
\label{eq:g1}
\end{equation}
\begin{equation}
\frac{\partial T_{s}}{\partial t}-\bar{V}\frac{\partial T_{s}}{\partial y}
=\kappa_{s}\left(\frac{\partial^{2} T_{s}}{\partial x^{2}}+\frac{\partial^{2} T_{s}}{\partial y^{2}}\right),
\label{eq:g2}
\end{equation}
where $t$ is time, $u$ and $v$ are the velocity components in the $x$ and $y$ direction measured in the laboratory frame in which the crystal is at rest, $\kappa_{l}$ and $\kappa_{s}$ are the thermal diffusivities of the liquid and solid, respectively. We substitute $T_{l}=\bar{T}_{l}+T'_{l}$, $T_{s}=\bar{T}_{s}+T'_{s}$, $u=\bar{U}+u'$ and $v=\Delta\rho\bar{V}+v'$ into Eqs. (\ref{eq:g1}) and (\ref{eq:g2}), where $\bar{T}_{l}$, $\bar{T}_{s}$ and $\bar{U}$ are unperturbed steady fields  and $T'_{l}$, $T'_{s}$, $u'$ and $v'$ are perturbed fields, respectively. Here $\Delta \rho \bar{V}$ is the advection flow due to the density difference of the liquid and solid, $\Delta \rho=(\rho_{l}-\rho_{s})/\rho_{l}$, $\rho_{l}$ and $\rho_{s}$ being the density of the liquid and solid. Then the equations for the unperturbed fields are\begin{equation}
\frac{d^{2} \bar{T}_{l}}{dy^{2}}+\frac{\rho \bar{V}}{\kappa_{l}}\frac{d \bar{T}_{l}}{dy}=0,
\label{eq:g3}
\end{equation}
\begin{equation}
\frac{d^{2} \bar{T}_{s}}{dy^{2}}+\frac{\bar{V}}{\kappa_{s}}\frac{d \bar{T}_{s}}{dy}=0,
\label{eq:g4}
\end{equation}
and the equations for the perturbed fields are
\begin{equation}
\frac{\partial T'_{l}}{\partial t}-\rho\bar{V}\frac{\partial T'_{l}}{\partial y}+\bar{U}\frac{\partial T'_{l}}{\partial x}+v'\frac{d \bar{T}_{l}}{dy}
=\kappa_{l}\left(\frac{\partial^{2} T'_{l}}{\partial x^{2}}+\frac{\partial^{2} T'_{l}}{\partial y^{2}}\right),
\label{eq:g5}
\end{equation}
\begin{equation}
\frac{\partial T'_{s}}{\partial t}-\rho\bar{V}\frac{\partial T'_{s}}{\partial y}=\kappa_{s}\left(\frac{\partial^{2} T'_{s}}{\partial x^{2}}+\frac{\partial^{2} T'_{s}}{\partial y^{2}}\right),
\label{eq:g6}
\end{equation}
where $\rho=\rho_{s}/\rho_{l}$.

Suppose that perturbations of the solid-liquid interface, temperature and normal flow are expressed in the following forms:
\begin{equation}
\zeta(t,x)=\zeta_{k} \exp[\sigma t+ikx],
\label{eq:g7}
\end{equation}
\begin{equation}
T'_{l}=g_{l}(y)\exp\left(-\frac{\rho \bar{V}}{2\kappa_{l}}y\right) \exp[\sigma t+ikx],
\label{eq:g8}
\end{equation}
\begin{equation}
T'_{s}=g_{s}(y)\exp\left(-\frac{\rho \bar{V}}{2\kappa_{s}}y\right) \exp[\sigma t+ikx],
\label{eq:g9}
\end{equation}
\begin{equation}
v'=v_{k} \exp[\sigma t+ikx],
\label{eq:g10}
\end{equation}
where $k$ is the wave number, and $\sigma=\sigma_{r}+i\sigma_{i}$, $\sigma_{r}$ being the rate of amplification or damping and $-\sigma_{i}/k$ being the phase velocity of the disturbance, $\zeta_{k}$ and $v_{k}$ are the amplitudes of perturbed interface and perturbed normal flow, $g_{l}$ and $g_{s}$ are the amplitudes of perturbed temperature of the liquid and solid, respectively. Substitutions of them into Eqs. (\ref{eq:g5}) and (\ref{eq:g6}) yield 
\begin{equation}
\frac{d^{2}g_{l}}{dy^{2}}-\left\{k^{2}+\left(\frac{\rho\bar{V}}{2\kappa_{l}}\right)^{2}+\frac{\sigma}{\kappa_{l}}+\frac{ik\bar{U}(y)}{\kappa_{l}}\right\}g_{l}=\frac{v_{k}}{\kappa_{l}}\frac{d \bar{T}_{l}}{dy} \exp\left(\frac{\rho \bar{V}}{2\kappa_{l}}y\right),
\label{eq:g11}
\end{equation}
\begin{equation}
\frac{d^{2}g_{s}}{dy^{2}}-\left\{k^{2}+\left(\frac{\rho\bar{V}}{2\kappa_{s}}\right)^{2}+\frac{\sigma}{\kappa_{s}}\right\}g_{s}=0.
\label{eq:g12}
\end{equation}

The following calculations are to first order only in the amplitude of the initial perturbation. The continuity of the temperature at the perturbed solid-liquid interface, $y=\zeta(t,x)$, is
\begin{equation}
(\bar{T}_{l}+T'_{l})|_{y=\zeta}=(\bar{T}_{s}+T'_{s})|_{y=\zeta}
=T_{m}+G(k)\zeta,
\label{eq:b1}
\end{equation}
where $T_{m}$ is the melting temperature, and $G(k)\zeta$ is the temperature difference from $T_{m}$ due to a deformation of the solid-liquid interface. The form of $G(k)$ will be specified later. Linearizing Eq. (\ref{eq:b1}) at $y=0$, Eq. (\ref{eq:b1}) gives to the zeroth order in $\zeta_{k}$, 
\begin{equation}
\bar{T}_{l}|_{y=0}=\bar{T}_{s}|_{y=0}=T_{m},
\label{eq:b2}
\end{equation}
and to the first order in $\zeta_{k}$,  
\begin{equation}
\frac{d \bar{T}_{l}}{dy}\Big|_{y=0}\zeta_{k}+g_{l}|_{y=0}
=\frac{d \bar{T}_{s}}{dy}\Big|_{y=0}\zeta_{k}+g_{s}|_{y=0}
=G(k)\zeta_{k}.
\label{eq:b3}
\end{equation}
It follows from Eq. (\ref{eq:b3}) that the amplitudes of $g_{l}|_{y=0}$ and $g_{s}|_{y=0}$ are of order $\zeta_{k}$:
\begin{equation}
g_{l}|_{y=0}=\left(-\frac{d \bar{T}_{l}}{dy}\Big|_{y=0}
+G(k)\right)\zeta_{k},
\label{eq:b4}
\end{equation}
\begin{equation}
g_{s}|_{y=0}=\left(-\frac{d \bar{T}_{s}}{dy}\Big|_{y=0}
+G(k)\right)\zeta_{k}.
\label{eq:b5}
\end{equation}

The heat conservation at the perturbed solid-liquid interface is
\begin{equation} 
L\left(\bar{V}+\frac{\partial \zeta}{\partial t} \right)
=K_{s}\frac{\partial (\bar{T}_{s}+T'_{s})}{\partial y}\Big|_{y=\zeta}
      -K_{l}\frac{\partial (\bar{T}_{l}+T'_{l})}{\partial y}\Big|_{y=\zeta},
\label{eq:b6}
\end{equation}
where $L$ is the latent heat per unit volume and $K_{s}$ and $K_{l}$ are the thermal conductivities of the solid and liquid, respectively. We linearize in the same way Eq. (\ref{eq:b6}) at $y=0$,  Eq. (\ref{eq:b6}) gives to the zeroth order in $\zeta_{k}$,
\begin{equation}
L\bar{V}=K_{s}\frac{d \bar{T}_{s}}{dy}\Big|_{y=0}
-K_{l}\frac{d\bar{T}_{l}}{dy}\Big|_{y=0},
\label{eq:b7}
\end{equation}
and to the first order in $\zeta_{k}$,
\begin{eqnarray}
L\sigma \zeta_{k}&=&
K_{s}\left(\frac{d^{2}\bar{T}_{s}}{dy^{2}}\Big|_{y=0}\zeta_{k}
-\frac{\rho \bar{V}}{2\kappa_{s}}g_{s}|_{y=0}
+\frac{dg_{s}}{dy}\Big|_{y=0}\right) \nonumber \\
& &-K_{l}\left(\frac{d^{2}\bar{T}_{l}}{dy^{2}}\Big|_{y=0}\zeta_{k}
-\frac{\rho \bar{V}}{2\kappa_{l}}g_{l}|_{y=0}
+\frac{dg_{l}}{dy}\Big|_{y=0}\right).
\label{eq:b8}
\end{eqnarray}
By substitutions of Eqs. (\ref{eq:b4}) and (\ref{eq:b5}) into Eq. (\ref{eq:b8}), the dispersion relation for the fluctuation of the solid-liquid interface becomes
\begin{eqnarray}
\sigma
&=&\frac{K_{s}}{L}\left\{\frac{d^{2}\bar{T}_{s}}{dy^{2}}\Big|_{y=0}
+\left(-\frac{\rho \bar{V}}{2\kappa_{s}}
+Q_{s}\right)\left(-\frac{d \bar{T}_{s}}{dy}\Big|_{y=0}
+G(k)\right) \right\} \nonumber \\
& & -\frac{K_{l}}{L}\left\{\frac{d^{2}\bar{T}_{l}}{dy^{2}}\Big|_{y=0}
+\left(-\frac{\rho \bar{V}}{2\kappa_{l}}
-Q_{l}\right)\left(-\frac{d \bar{T}_{l}}{dy}\Big|_{y=0}
+G(k)\right)\right\} ,
\label{eq:b9}
\end{eqnarray}
where we have defined the so-called propagator in the liquid and solid as follows \cite{5}:
\begin{equation}
Q_{l}=-\frac{\frac{dg_{l}}{dy}\Big|_{y=0}}{g_{l}|_{y=0}}, 
\hspace{1cm}
Q_{s}=\frac{\frac{dg_{s}}{dy}\Big|_{y=0}}{g_{s}|_{y=0}}, 
\label{eq:b10}
\end{equation}
which describe the motion of the interface in response to the propagation of a temperature disturbance, here it is the latent heat release. The general formulation above will be applied in Sec. V.

\section{The perturbed normal flow over the sinusoidal interface}
It is firstly necessary to know the primary shear flow field $\bar{U}(y)$ and the amplitude $v_{k}$ of the perturbed normal flow in Eq. (\ref{eq:g11}). 
We  determine the perturbed normal flow over the interface to first order only in the amplitude of initial perturbation by following the formulation of Benjamin \cite{30}. In his treatment, the bottom is flat interface and does not move. When the crystal grows, however, the solid-liquid interface moves and it may be not flat if a morphological instability occurs. Therefore, it is necessary to modify several boundary conditions in his formulation. We make use of the same dimensionless variables as those used in Benjamin's paper, which are different from those used in Ogawa-Furukawa's paper \cite{20}. Hereafter we refer to their models as O-F model.

With reference to Fig. \ref{fig:diagram}, the primary shear flow assumed steady, is pararell to the $x$ axis, with the velocity $\bar{U}$ varing only with $y$. In the frame of reference moving at the solid-liquid interface velocity $\bar{V}$, the Navier-Stokes equations are
\begin{equation}
\frac{\partial u}{\partial t}-\bar{V}\frac{\partial u}{\partial y}+u\frac{\partial u}{\partial x}+v\frac{\partial u}{\partial y}
=-\frac{1}{\rho_{l}}\frac{\partial p}{\partial x}
+\nu\left(\frac{\partial^{2}u}{\partial x^{2}}+\frac{\partial^{2}u}{\partial y^{2}}\right)+g \sin\theta,
\label{eq:f1} 
\end{equation}
\begin{equation}
\frac{\partial v}{\partial t}-\bar{V}\frac{\partial v}{\partial y}+u\frac{\partial v}{\partial x}+v\frac{\partial v}{\partial y}
=-\frac{1}{\rho_{l}}\frac{\partial p}{\partial y}
+\nu\left(\frac{\partial^{2}v}{\partial x^{2}}+\frac{\partial^{2}v}{\partial y^{2}}\right)-g \cos\theta, 
\label{eq:f2}
\end{equation}
where $t$ is time, $u$ ($v$) the velocity components in the $x$ $(y)$ direction, $p$ the pressure, $\rho_{l}$ the liquid density, $g$ the gravitational acceleration, $\nu$ the kinematic viscosity, and $\theta$ the angle of the inclined plane. The equation of continuity is
\begin{equation}
\frac{\partial u}{\partial x}+\frac{\partial v}{\partial y}=0.
\label{eq:f3}
\end{equation}

In this section the coordinates $(x,y)$ and velocities $(u,v)$ are made non-dimensional by taking the mean thickness $h_{0}$ of flowing liquid as the unit of length and the velocity $u_{0}$ at the free surface as the unit of velocity respectively. By substitutions of $(x_{*},y_{*})=(x,y)/h_{0}$, $(u_{*},v_{*})=(u,v)/u_{0}$, $p_{*}=p/(\rho_{l} u_{0}^{2})$ and $t_{*}=tu_{0}/h_{0}$, the equations of motion and continuity can be written in the following dimensionless forms:
\begin{equation}
\frac{\partial u_{*}}{\partial t_{*}}-\bar{V}_{*}\frac{\partial u_{*}}{\partial y_{*}}+u_{*}\frac{\partial u_{*}}{\partial x_{*}}+v_{*}\frac{\partial u_{*}}{\partial y_{*}}=-\frac{\partial p_{*}}{\partial x_{*}}
+\frac{1}{\rm Re}\left(\frac{\partial^{2}u_{*}}{\partial x_{*}^{2}}+\frac{\partial^{2}u_{*}}{\partial y_{*}^{2}}\right)+\frac{\sin\theta}{\rm F^{2}}, 
\label{eq:f4}
\end{equation}
\begin{equation}
\frac{\partial v_{*}}{\partial t_{*}}-\bar{V}_{*}\frac{\partial v_{*}}{\partial y_{*}}+u_{*}\frac{\partial v_{*}}{\partial x_{*}}+v_{*}\frac{\partial v_{*}}{\partial y_{*}}=-\frac{\partial p_{*}}{\partial y_{*}}
+\frac{1}{\rm Re}\left(\frac{\partial^{2}v_{*}}{\partial x_{*}^{2}}+\frac{\partial^{2}v_{*}}{\partial y_{*}^{2}}\right)-\frac{\cos\theta}{\rm F^{2}}, 
\label{eq:f5}
\end{equation}
\begin{equation}
\frac{\partial u_{*}}{\partial x_{*}}+\frac{\partial v_{*}}{\partial y_{*}}=0,
\label{eq:f6}
\end{equation}
where ${\rm Re}=u_{0}h_{0}/\nu$ is the Reynolds number and ${\rm F}=u_{0}/(gh_{0})^{1/2}$ is the Froude number. 

Let
\begin{equation}
u_{*}=\bar{U}_{*}+u'_{*}, \hspace{5mm} 
v_{*}=\Delta\rho\bar{V}_{*}+ v'_{*}, \hspace{5mm} 
p_{*}=\bar{P}_{*}+p'_{*},
\label{eq:f7}
\end{equation}
where $\bar{U}_{*}$ and $\bar{P}_{*}$ are the dimensionless velocity and pressure of the primary flow, and primed quantities are the dimensionless velocity and pressure perturbations.
Substitution of Eq. (\ref{eq:f7}) into Eqs. (\ref{eq:f4}), (\ref{eq:f5}) and (\ref{eq:f6}) yields
\begin{equation}
\frac{1}{\rm Re}\frac{d^{2}\bar{U}_{*}}{dy_{*}^{2}}+\rho\bar{V}_{*}\frac{d\bar{U}_{*}}{dy_{*}}+\frac{1}{\rm F^{2}} \sin\theta=0,
\label{eq:f8}
\end{equation}
\begin{equation}
\frac{d\bar{P}_{*}}{dy}+\frac{1}{\rm F^{2}} \cos\theta=0,
\label{eq:f9}
\end{equation}
\begin{equation}
\frac{\partial u'_{*}}{\partial t_{*}}-\rho\bar{V}_{*}\frac{\partial u'_{*}}{\partial y_{*}}+\bar{U}_{*}\frac{\partial u'_{*}}{\partial x_{*}}+\frac{d\bar{U}_{*}}{dy_{*}}v'_{*}
=-\frac{\partial p'_{*}}{\partial x_{*}}+\frac{1}{\rm Re}\left(\frac{\partial^{2} u'_{*}}{\partial x_{*}^{2}}+\frac{\partial^{2} u'_{*}}{\partial y_{*}^{2}}\right),\label{eq:f10}
\end{equation}
\begin{equation}
\frac{\partial v'_{*}}{\partial t_{*}}-\rho\bar{V}_{*}\frac{\partial v'_{*}}{\partial y_{*}}+\bar{U}_{*}\frac{\partial v'_{*}}{\partial x_{*}}
=-\frac{\partial p'_{*}}{\partial y_{*}}+\frac{1}{\rm Re}\left(\frac{\partial^{2} v'_{*}}{\partial x_{*}^{2}}+\frac{\partial^{2} v'_{*}}{\partial y_{*}^{2}}\right),\label{eq:f11}
\end{equation}
\begin{equation}
\frac{\partial u'_{*}}{\partial x_{*}}+\frac{\partial v'_{*}}{\partial y_{*}}=0,\label{eq:f12}
\end{equation}
if quadratic terms in the perturbation quantities are neglected. Using the typical values of $\bar{V} \sim 10^{-6}$ m/s and $u_{0} \sim 10^{-2}$ m/s in the experiments \cite{21,22}, we can neglect the $\rho\bar{V}_{*}$ term in Eqs. (\ref{eq:f8}), (\ref{eq:f10}) and (\ref{eq:f11}) because $\bar{V}_{*}$ is the ratio of $\bar{V}$ to $u_{0}$. 

Under the boundary conditions, 
\begin{equation}
\bar{U}_{*}=0 \hspace{5mm} (y_{*}=0), \hspace{1cm} 
\frac{d\bar{U}_{*}}{dy_{*}}=0 \hspace{5mm} (y_{*}=1), \hspace{1cm} 
\bar{P}_{*}=P_{0*} \hspace{5mm} (y_{*}=1),
\label{eq:f13}
\end{equation}
the solutions of Eqs. (\ref{eq:f8}) and (\ref{eq:f9}) are respectively \cite{31},
\begin{equation}
\bar{U}_{*}=2y_{*}-y_{*}^{2},
\label{eq:f14}
\end{equation}
\begin{equation}
\bar{P}_{*}=P_{0*}+\frac{\cos\theta}{\rm F^{2}}(1-y_{*}),
\label{eq:f15}
\end{equation}
where $P_{0*}$ is the dimensionless pressure of atmosphere.
Eq. (\ref{eq:f12}) allows the use of a stream function $\psi'$, in terms of which $u'_{*}$ and $v'_{*}$ can be expressed as follows: 
\begin{equation}
u'_{*}=\frac{\partial \psi'}{\partial y_{*}},\hspace{1cm} 
v'_{*}=-\frac{\partial \psi'}{\partial x_{*}}.
\label{eq:f16}
\end{equation}
Eqs. (\ref{eq:f10}) and (\ref{eq:f11}) can then be written as 
\begin{equation}
\frac{\partial^{2} \psi'}{\partial t_{*}\partial y_{*}}+\bar{U}_{*}\frac{\partial^{2} \psi'}{\partial x_{*}\partial y_{*}}-\frac{d\bar{U}_{*}}{dy_{*}}\frac{\partial \psi'}{\partial x_{*}}
=-\frac{\partial p'_{*}}{\partial x_{*}}+\frac{1}{\rm Re}\left(\frac{\partial^{3} \psi'}{\partial x_{*}^{2}\partial y_{*}}+\frac{\partial^{3} \psi'}{\partial y_{*}^{3}}\right),
\label{eq:f17}
\end{equation}
\begin{equation}
\frac{\partial^{2} \psi'}{\partial t_{*}\partial x_{*}}+\bar{U}_{*}\frac{\partial^{2} \psi'}{\partial x_{*}^{2}}
=\frac{\partial p'_{*}}{\partial y_{*}}+\frac{1}{\rm Re}\left(\frac{\partial^{3} \psi'}{\partial x_{*}^{3}}+\frac{\partial^{3} \psi'}{\partial x_{*}\partial y_{*}^{2}}\right).
\label{eq:f18}
\end{equation}

If the perturbation of the solid-liquid interface is represented in a dimensionless form, 
\begin{equation}
\zeta_{*}(t_{*},x_{*})=\delta_{b} \exp[\sigma_{*}t_{*}+i\mu x_{*}],
\label{eq:f19}
\end{equation}
the corresponding perturbations of the stream function, pressure and the liquid-air surface may be written as respectively, 
\begin{equation}
\psi'=\delta_{b}f(y_{*}) \exp[\sigma_{*}t_{*}+i\mu x_{*}],
\label{eq:f20}
\end{equation}
\begin{equation}
p'_{*}=\delta_{b}\Pi(y_{*}) \exp[\sigma_{*}t_{*}+i\mu x_{*}],
\label{eq:f21}
\end{equation}
\begin{equation}
\xi_{*}(t_{*},x_{*})=1+\delta_{t} \exp[\sigma_{*}t_{*}+i\mu x_{*}],
\label{eq:f22}
\end{equation}
in which $\delta_{b}=\zeta_{k}/h_{0}$ and $\delta_{t}=\xi_{k}/h_{0}$ are dimensionless amplitudes of the solid-liquid interface and the liquid-air surface respectively, $\mu=kh_{0}$ is the dimensionless wave number, and $\sigma_{*}=\sigma h_{0}/u_{0}$. When we substitute Eqs. (\ref{eq:f20}) and (\ref{eq:f21}) into Eqs. (\ref{eq:f17}) and (\ref{eq:f18}), and  $\Pi$ is eliminated from them by cross differentiation, the linearized equations of motion lead to the Orr-Sommerfeld equation:
\begin{eqnarray}
\lefteqn{\frac{d^{4}f}{dy_{*}^{4}}-2\mu^{2}\frac{d^{2}f}{dy_{*}^{2}}+\mu^{4}f} \hspace*{1cm} \nonumber \\
& = & i\mu {\rm Re}\left\{\left(\bar{U}_{*}-i\frac{\sigma_{*}}{\mu}\right)\left(\frac{d^{2}f}{dy_{*}^{2}}-\mu^{2}f\right)
-\frac{d^{2}\bar{U}_{*}}{dy_{*}^{2}}f\right\}.
\label{eq:f23}
\end{eqnarray}
The perturbed flow was assumed to be stationary from the outset in the O-F model. This formally amounts to neglecting the $\sigma_{*}/\mu$ term in Eq. (\ref{eq:f23}). This assumption will be justified in Sec. VI. Since the value of the mean thickness $h_{0}$ is about $10^{-4}$ m, and the typical value of wavelength of the wavy pattern observed on the surface of icicles is about 1 cm \cite{21,22}, the value of $\mu=kh_{0}$ is about $6\times 10^{-2}$ \cite{20}. Therefore, in the long wavelength approximation retaining up to the first order in $\mu$, Eq. (\ref{eq:f23}) becomes
\begin{equation}
\frac{d^{4}f}{dy_{*}^{4}}
=i\mu {\rm Re}\left\{(2y_{*}-y_{*}^{2})\frac{d^{2}f}{dy_{*}^{2}}+2f\right\},
\label{eq:f24}
\end{equation}
where we have substituted Eq. (\ref{eq:f14}) for $\bar{U}_{*}$. We note that Re becomes $O(1)$ when we use the typical values of $u_{0}$ and $h_{0}$ used above, and $\nu=1.8 \times 10^{-6}$ $\rm m^{2}/s$ of water, therefore the primary shear flow is laminar \cite{20,21,22}.

The problem entails five boundary conditions as follows. 
Since both velocity components must vanish at the perturbed solid-liquid interface, we have
\begin{equation}
v'_{*}|_{y_{*}=\zeta_{*}}
-\Delta \rho\frac{\partial \zeta_{*}}{\partial t_{*}}=0,
\label{eq:fb1}
\end{equation}
\begin{equation}
(\bar{U}_{*}+u'_{*})|_{y_{*}=\zeta_{*}}
+\Delta\rho\bar{V}_{*}\frac{\partial \zeta_{*}}{\partial x_{*}}=0.
 \label{eq:fb2}
\end{equation}
The kinematic condition at the free surface is
\begin{equation}
\frac{\partial \xi_{*}}{\partial t_{*}}+\bar{U}_{*}\frac{\partial \xi_{*}}{\partial x_{*}}=v'_{*}|_{y_{*}=\xi_{*}}.
\label{eq:fb3}
\end{equation}
At the free surface the shear stress must vanish and the normal stress must just balance the normal stress induced by surface tension; 
\begin{equation}
\frac{\partial u_{*}}{\partial y_{*}}\Big|_{y_{*}=\xi_{*}}
+\frac{\partial v_{*}}{\partial x_{*}}\Big|_{y_{*}=\xi_{*}}=0,
\label{eq:fb4}
\end{equation}
\begin{equation}
-p_{*}|_{y_{*}=\xi_{*}}+\frac{2}{\rm Re}\frac{\partial v_{*}}{\partial y_{*}}\Big|_{y_{*}=\xi_{*}}-{\rm S}\frac{\partial^{2}\xi_{*}}{\partial x_{*}^2}=0,
\label{eq:fb5}
\end{equation}
where S=$\gamma/(\rho_{l} h_{0}u_{0}^2)$, $\gamma$ being the surface tension of the liquid-air.
Linearizing Eqs. (\ref{eq:fb1}) and (\ref{eq:fb2}) at $y_{*}=0$ and Eqs. (\ref{eq:fb3}), (\ref{eq:fb4}) and (\ref{eq:fb5}) at $y_{*}=1$, Eqs. (\ref{eq:fb1}) to (\ref{eq:fb5}) become respectively,
\begin{equation}
f|_{y_{*}=0}=i\Delta \rho\frac{\sigma_{*}}{\mu},
\label{eq:fb6}
\end{equation}
\begin{equation}
\frac{df}{dy_{*}}\Big|_{y_{*}=0}=-2-i\mu\Delta\rho\bar{V}_{*}, 
\label{eq:fb7}
\end{equation}
\begin{equation}
f|_{y_{*}=1}\delta_{b}=\left(i\frac{\sigma_{*}}{\mu}-1\right)\delta_{t},
\label{eq:fb8}
\end{equation}
\begin{equation}
\left(\frac{d^{2}f}{dy_{*}^{2}}\Big|_{y_{*}=1}+\mu^{2}f|_{y_{*}=1}\right)
\delta_{b}=2\delta_{t},
\label{eq:fb9}
\end{equation}
\begin{eqnarray}
\lefteqn{\frac{d^{3}f}{dy_{*}^{3}}\Big|_{y_{*}=1}\delta_{b}-i\left(\frac{\mu {\rm Re} \cos\theta}{\rm F^{2}}+\mu^{3}\rm ReS\right)\delta_{t}} \hspace*{2cm} \nonumber \\&=&\left\{i\mu {\rm Re}\left(1-i\frac{\sigma_{*}}{\mu}\right)+3\mu^{2}\right\}\frac{df}{dy_{*}}\Big|_{y_{*}=1}\delta_{b}
-\rho\bar{V_{*}}{\rm Re}\frac{d^{2}f}{dy_{*}^{2}}\Big|_{y_{*}=1}\delta_{b}.
\label{eq:fb10}
\end{eqnarray}


If we formaly put
\begin{equation}
f(y_{*})=\sum_{N=0}^{\infty}A_{N}y_{*}^{N},
\label{eq:se1}
\end{equation}
then this series is seen to constitute a solution of Eq. (\ref{eq:f24}) when the coefficients $A_{N}$ are made to satisfy the following recursion relation
\begin{eqnarray}
\lefteqn{
N(N-1)(N-2)(N-3)A_{N}} \hspace*{2cm} \nonumber \\
&=&2i\mu {\rm Re}(N-3)(N-4)A_{N-3}+i\mu {\rm Re}\{2-(N-4)(N-5)\}A_{N-4},
\label{eq:se2}
\end{eqnarray}
for $N>3$. 
Eq. (\ref{eq:se2}) gives every other $A_{N}$ in terms of the first four coefficients $A_{0}$ to $A_{3}$. The approximation to the series solution up to the first order in $\mu$ requires seven coefficients of the expansion Eq. (\ref{eq:se1}). Therefore, the other coefficients are given as follows:
\begin{equation}
A_{4}=0, 
\label{eq:se3}
\end{equation}
\begin{equation}
A_{5}=\frac{i\mu {\rm Re}}{60}A_{1}+\frac{i\mu {\rm Re}}{30}A_{2}, 
\label{eq:se4}
\end{equation}
\begin{equation}
A_{6}=\frac{i\mu {\rm Re}}{30}A_{3}, 
\label{eq:se5}
\end{equation}
\begin{equation}
A_{7}=-\frac{i\mu {\rm Re}}{210}A_{3}.
\label{eq:se6}
\end{equation}
Hence, the approximate series solution can be written as
\begin{eqnarray}
\lefteqn{f(y_{*})
=\left(1+\frac{i\mu {\rm Re}}{12}y_{*}^{4}\right)A_{0} 
+\left(y_{*}+\frac{i\mu {\rm Re}}{60}y_{*}^{5}\right)A_{1}}\hspace{1cm} \nonumber \\
& &+\left(y_{*}^{2}+\frac{i\mu {\rm Re}}{30}y_{*}^{5}\right)A_{2}
+\left(y_{*}^{3}+\frac{i\mu {\rm Re}}{30}y_{*}^{6}
-\frac{i\mu {\rm Re}}{210}y_{*}^{7}\right)A_{3}. 
\label{eq:se7}
\end{eqnarray}
The four constants $A_{0}$ to $A_{3}$ of the solution of the fourth order Eq. (\ref{eq:f24}) are determined from the boundary conditions Eqs. (\ref{eq:fb6}) to (\ref{eq:fb10}) in the form neglecting the terms including $\bar{V}_{*}$, $\sigma_{*}/\mu$ and $\mu^{2}$.

First, the boundary conditions Eqs. (\ref{eq:fb6}) and (\ref{eq:fb7}) give respectively,
\begin{equation}
A_{0}=0, 
\label{eq:se8}
\end{equation}
\begin{equation}
A_{1}=-2. 
\label{eq:se9}
\end{equation}
Eliminations of $\delta_{t}$ from Eqs. (\ref{eq:fb8}) and (\ref{eq:fb9}) and from Eqs. (\ref{eq:fb8}) and (\ref{eq:fb10}) yield respectively,
\begin{equation}
\frac{d^{2}f}{dy_{*}^{2}}\Big|_{y_{*}=1}=-2f|_{y_{*}=1},
\label{eq:se10}
\end{equation}
\begin{equation}
\frac{d^{3}f}{dy_{*}^{3}}\Big|_{y_{*}=1}+i\alpha f|_{y_{*}=1} 
=i\mu {\rm Re}\frac{df}{dy_{*}}\Big|_{y_{*}=1},
\label{eq:se11}
\end{equation}
where 
\begin{equation}
\alpha=\frac{\mu {\rm Re} \cos\theta}{\rm F^{2}}+\mu^{3}{\rm Re S}
=\frac{gh_{0}^{3} \cos\theta}{\nu u_{0}}k
+\frac{\gamma h_{0}^{3}}{\rho_{l}\nu u_{0}}k^{3}.
\label{eq:se12}
\end{equation}
Eq. (\ref{eq:se12}) represents the restoring forces due to gravity and surface tension \cite{30,31}. When we use the typical values of $u_{0}\sim10^{-2}$ m/s and $h_{0}\sim10^{-4}$ m in the experiments \cite{21,22}, and the physical properties of water, $\rho_{l}=1.0 \times 10^{3}$ $\rm Kg/m^{3}$, $\nu=1.8 \times 10^{-6}$ $\rm m^{2}/s$ and $\gamma=7.6\times 10^{-2}$ N/m, $\alpha$ becomes $O(1)$ for the wavelength of wavy pattern occuring on the icicles and the inclined plane, therefore we treat $\alpha$ as zeroth order in terms of $\mu$ in the following calculations. Since $u_{0}$ and $h_{0}$ are not independent quantities, this rough order estimate will be justified more quantitatively in Sec. VI by using other parameters being controlled easily in the actual experiment.

Using Eqs. (\ref{eq:se7}), (\ref{eq:se8}) and (\ref{eq:se9}), Eqs. (\ref{eq:se10}) and (\ref{eq:se11}) give respectively,
\begin{equation}
\left(4+\frac{11i\mu {\rm Re}}{15}\right)A_{2}+\left(8+\frac{6i\mu {\rm Re}}{7}\right)A_{3}
=4+\frac{11i\mu {\rm Re}}{15},
\label{eq:se13}
\end{equation}
\begin{equation}
i\alpha\left(1+\frac{i\mu {\rm Re}}{30}\right)A_{2}+\left(i\alpha+6-\frac{\mu {\rm Re}\alpha}{35}\right)A_{3}
=i\alpha\left(2+\frac{i\mu {\rm Re}}{30}\right).
\label{eq:se14}
\end{equation}
Retaining up to the first order in $\mu$, the solutions of these simultaneous equations for $A_{2}$ and $A_{3}$ are expressed as follows: 
\begin{equation}
A_{2}=\frac{3(2-i\alpha)}{6-i\alpha}
+\mu {\rm Re}\alpha\frac{-96-8i\alpha}{105(6-i\alpha)^{2}},
\label
{eq:se15}
\end{equation}
\begin{equation}
A_{3}=\frac{i\alpha}{6-i\alpha}
+\mu {\rm Re}\alpha\frac{4i\alpha}{35(6-i\alpha)^{2}}.
\label{eq:se16}
\end{equation}
When these expressions of $A_{0}$ to $A_{3}$ are substituted into Eq. (\ref{eq:se7}), the final form up to the first order in $\mu$ is
\begin{eqnarray}
f(y_{*})&=&
-2y_{*}+\frac{3(2-i\alpha)}{6-i\alpha}y_{*}^{2}
+\frac{i\alpha}{6-i\alpha}y_{*}^{3}
+\mu {\rm Re}\alpha\left\{\frac{-96-8i\alpha}{105(6-i\alpha)^{2}}y_{*}^{2}
      +\frac{4i\alpha}{35(6-i\alpha)^{2}}y_{*}^{3}\right.
       \nonumber \\
& & \left.+\frac{1}{15(6-i\alpha)}y_{*}^{5}
          -\frac{1}{30(6-i\alpha)}y_{*}^{6}
        +\frac{1}{210(6-i\alpha)}y_{*}^{7}\right\}.
\label{eq:se17}
\end{eqnarray}
Applying this result at $y_{*}=1$ to Eq. (\ref{eq:fb8}), we can know the relation between the amplitude and the phase of perturbation of the solid-liquid interface and that of the liquid-air surface:
\begin{equation}
\delta_{t}=-f|_{y_{*}=1}\delta_{b}.
\label{eq:se18}
\end{equation}
In the article \cite{20}, the following function was obtained: 
\begin{equation}
f(y_{*})=-2y_{*}+y_{*}^{2}.
\label{eq:se19}
\end{equation}
If we substitute Eq. (\ref{eq:se19}) into Eq. (\ref{eq:se18}), $\delta_{t}=\delta_{b}$, which indicates that the liquid-air surface fluctuates with the same amplitude as the solid-liquid interface and phase shift of each interface does not occur. If we regard $\alpha$ as $O(1)$ with respect to $\mu$, however, Eq. (\ref{eq:se19}) can not satisfy the boundary condition Eq. (\ref{eq:se11}). On the other hand, if we substitute Eq. (\ref{eq:se17}) into Eq. (\ref{eq:se18}), it is found that the amplitude and the phase of the liquid-air surface depends on the wavelength of a fluctuation of the solid-liquid interface because of the restoring forces $\alpha$.

By rewriting the second equation of Eq. (\ref{eq:f16}) in the dimensional form,
\begin{equation}
v'=-u_{0}h_{0}\frac{\partial \psi'}{\partial x}=-iku_{0}f(y)\zeta_{k}
 \exp[\sigma t+ikx],
\label{eq:se20}
\end{equation}
and by comparing it with Eq. (\ref{eq:g10}) we obtain
\begin{equation}
v_{k}=-iku_{0}f(y)\zeta_{k},
\label{eq:se21}
\end{equation}
where $f(y)$ is given by Eq. (\ref{eq:se17}) in the long wavelength approximation retaining up to the first order in $\mu$ .

\section{General solution for the perturbed temperature distribution in the liquid}
In the preceding section, we have determined $\bar{U}(y)$ and $v_{k}$ in Eq. (\ref{eq:g11}). Next, we must determine the amplitude of the perturbed temperature in the liquid under this primary shear flow and the amplitude of perturbed normal flow. 
Since the Peclet number $\bar{V}h_{0}/\kappa_{l}$ associated with the crystal growth velocity is very small when the typical values of $\bar{V}\sim10^{-6}$ ${\rm m/s}$, $h_{0}\sim10^{-4}$ m in the experiments \cite{21,22}, and $\kappa_{l}=1.3 \times 10^{-7}$ ${\rm m^{2}/s}$ of water are used, we can neglect the second term of Eq. (\ref{eq:g3}), then the solution is
\begin{equation}
\bar{T}_{l}(y)=T_{m}-\bar{G}_{l}y,
\label{eq:pt1}
\end{equation}
where $\bar{G}_{l}=(T_{m}-T_{la})/h_{0}$ is unperturbed temperature gradient in the liquid, $T_{la}$ is the temperature of the liquid-air surface. If we make the substitutions of $y=h_{0}(1-z)$, $\mu=kh_{0}$ and $u_{0}h_{0}/\kappa_{l} \equiv$ Pe, which is the Peclet number associated with the flow velocity at the free surface, into Eq. (\ref{eq:g11}), we obtain
\begin{eqnarray}
\lefteqn{
\frac{d^{2}g_{l}}{dz^{2}}-\left\{\mu^{2}
+\left(\frac{\rho\bar{V}h_{0}}{2\kappa_{l}}\right)^{2}
+\frac{\sigma h_{0}^{2}}{\kappa_{l}}+i\mu {\rm Pe} \right\}g_{l}
+i\mu {\rm Pe} z^{2}g_{l} 
} \hspace*{2cm} \nonumber \\
& = &
i\mu {\rm Pe}f(z) \exp\left\{-\frac{\rho \bar{V}h_{0}}{2\kappa_{l}}(1-z)\right\}\bar{G}_{l}\zeta_{k},
\label{eq:pt2}
\end{eqnarray}
where we have used Eqs. (\ref{eq:f14}), (\ref{eq:se21}) and (\ref{eq:pt1}).
When we put the right hand side of Eq. (\ref{eq:pt2}) equal to zero, Eq. (\ref{eq:pt2}) becomes to the equation  for a parabolic cylinder function, 
\begin{equation}
\frac{d^{2}\phi}{dz^{2}}-\left\{\mu^{2}
+\left(\frac{\rho\bar{V}h_{0}}{2\kappa_{l}}\right)^{2}
+\frac{\sigma h_{0}^{2}}{\kappa_{l}}+i\mu {\rm Pe} \right\}\phi+i\mu {\rm Pe} z^{2}\phi=0.
\label{eq:pt3}
\end{equation}

Using the confluent hypergeometric function $_{1}F_{1}$, the general solutions 
of Eq. (\ref{eq:pt3}) are given by \cite{40}
\begin{equation}
\phi_{1}(z)=\exp\left(-\frac{1}{2}(-i\mu {\rm Pe})^{1/2}z^2\right)\,\!_{1}F_{1}\left(\frac{1}{4}\left\{1+\frac{\mu^{2}+i\mu {\rm Pe}}{(-i\mu {\rm Pe})^{1/2}}\right\},\frac{1}{2},(-i\mu {\rm Pe})^{1/2}z^2 \right),
\label{eq:pt4}
\end{equation}
\begin{equation}
\phi_{2}(z)=z\exp\left(-\frac{1}{2}(-i\mu {\rm Pe})^{1/2}z^2\right)\,\!_{1}F_{1}\left(\frac{1}{2}+\frac{1}{4}\left\{1+\frac{\mu^{2}+i\mu {\rm Pe}}{(-i\mu {\rm Pe})^{1/2}}\right\},\frac{3}{2},(-i\mu {\rm Pe})^{1/2}z^2 \right).
\label{eq:pt5}
\end{equation}
Then we can show that the Wronskian $W$ of the two solutions $\phi_{1}(z)$ and $\phi_{2}(z)$ becomes 
\begin{equation}
W(z)=\phi_{1}(z)\frac{d\phi_{2}(z)}{dz}-\phi_{2}(z)\frac{d\phi_{1}(z)}{dz}=1.
\label{eq:pt6}
\end{equation}
Therefore, the solution of Eq. (\ref{eq:pt2}) is given as follows:
\begin{equation}
g_{l}(z)=B_{1}\phi_{1}(z)+B_{2}\phi_{2}(z)+i\mu {\rm Pe} \int_{0}^{z}\left\{\phi_{2}(z)\phi_{1}(z')-\phi_{1}(z)\phi_{2}(z')\right\}
f(z')dz'\bar{G}_{l}\zeta_{k},
\label{eq:pt7}
\end{equation}
where $B_{1}$ and $B_{2}$ are constants with respect to $z$, and  in Eq. (\ref{eq:pt7}) we have omitted the exponential term on the right hand side of Eq. (\ref{eq:pt2}) because $\bar{V}h_{0}/\kappa_{l} \ll 1$. In  Eqs. (\ref{eq:pt4}) and (\ref{eq:pt5}), we have omitted the terms $(\rho \bar{V}h_{0}/2\kappa_{l})^{2}$ and $\sigma h_{0}^{2}/\kappa_{l}$ in Eq. (\ref{eq:pt3}) because we can evaluate the ratio of the second term to the first one, $(\rho\bar{V}h_{0}/2\kappa_{l})^{2}/\mu^{2}=(\rho\bar{V}/2\kappa_{l}k)^{2} \ll 1$, and the ratio of the third term to the first one, $\sigma h_{0}^{2}/(\kappa_{l}\mu^{2})=\sigma/(\kappa_{l}k^{2}) \ll 1$. We are concerned with the wave number region that satisfy the former condition. While the latter condition amounts to neglecting the time dependence of the perturbed temperature field. It was assumed from the outset in the O-F model. This will be justified in Sec. VI. The constants $B_{1}$ and $B_{2}$ must be determined from the boundary conditions at the liquid-air surface. 

The equation for the temperature distribution in the air is 
\begin{equation}
\frac{\partial (\bar{T}_{a}+T'_{a})}{\partial t}
-\bar{V}\frac{\partial (\bar{T}_{a}+T'_{a})}{\partial y}
=\kappa_{a}\left(\frac{\partial^{2} }{\partial x^{2}}+\frac{\partial^{2} }{\partial y^{2}}\right)(\bar{T}_{a}+T'_{a}),
\label{eq:ab1}
\end{equation}
where $\bar{T}_{a}$ and $T'_{a}$ are unperturbed temperature and perturbed temperature of the air respectively, and $\kappa_{a}$ is the thermal diffusivity of the air. The solution for the unperturbed temperature field is
\begin{equation}
\bar{T}_{a}(y)=T_{\infty}+(T_{la}-T_{\infty}) \exp\left(-\frac{y-h_{0}}{l_{a}}\right),
\label{eq:ab2}
\end{equation}
where $T_{\infty}$ is the temperature of the air at $y=\infty$ and $l_{a}=\kappa_{a}/\bar{V}$ is the thermal diffusion length of the air. Suppose that the perturbed temperature distribution of the air is expressed in the following form:
\begin{equation}
T'_{a}=g_{a}(y) \exp\left[-\frac{\bar{V}}{2\kappa_{a}}(y-h_{0})\right] \exp[\sigma t+ikx],\label{eq:ab3}
\end{equation}
where
\begin{equation}
g_{a}(y)=T_{ka} \exp[-q(y-h_{0})],
\label{eq:ab4}
\end{equation}
and $T_{ka}$ is the amplitude of the perturbed temperature of the air.
Substitution of Eq. (\ref{eq:ab3}) into Eq. (\ref{eq:ab1}) gives
\begin{equation}
q=\sqrt{k^{2}+\left(\frac{\bar{V}}{2\kappa_{a}}\right)^{2}+\frac{\sigma}{\kappa_{a}}}.
\label{eq:ab5}
\end{equation}
In the quasi-stationary approximation $\sigma/(\kappa_{a}k^{2})\ll 1$ and $kl_{a} \gg 1$, we can approximate $q\cong k$. 

The continuity of the temperature at the liquid-air surface, $y=\xi(t,x)=h_{0}+\xi_{k}\exp[\sigma t+ikx]$, is 
\begin{equation}
(\bar{T}_{l}+T'_{l})|_{y=\xi}=(\bar{T}_{a}+T'_{a})|_{y=\xi}=T_{la}.
\label{eq:ab6}
\end{equation}
Linearizing Eq. (\ref{eq:ab6}) at $y=h_{0}$, Eq. (\ref{eq:ab6}) gives to the zeroth order in $\xi_{k}$,
\begin{equation}
\bar{T}_{l}|_{y=h_{0}}=\bar{T}_{a}|_{y=h_{0}}=T_{la},
\label{eq:ab7}
\end{equation}
and to the first order in $\xi_{k}$, 
\begin{equation}
-\bar{G}_{l}\xi_{k}+g_{l}|_{y=h_{0}}
 \exp\left(-\frac{\rho \bar{V}h_{0}}{2\kappa_{l}}\right)
=-\bar{G}_{a}\xi_{k}+T_{ka}=0,
\label{eq:ab8}
\end{equation}
where $\bar{G}_{a}=(T_{la}-T_{\infty})/l_{a}$.
Hereafter we omit the terms including $\bar{V}h_{0}/\kappa_{l}$ because $\bar{V}h_{0}/\kappa_{l} \ll 1$.
Heat conservation at the liquid-air surface is
\begin{equation}
-K_{l}\frac{\partial (\bar{T}_{l}+T'_{l})}{\partial y}\Big|_{y=\xi}=-K_{a}\frac{\partial (\bar{T}_{a}+T'_{a})}{\partial y}\Big|_{y=\xi},
\label{eq:ab9}
\end{equation}
where $K_{a}$ is the thermal conductivity of the air.
Similarly, linearizing Eq. (\ref{eq:ab9}) at $y=h_{0}$, Eq. (\ref{eq:ab9}) gives to the zeroth order in $\xi_{k}$,
\begin{equation}
K_{l}\bar{G}_{l}=K_{a}\bar{G}_{a},
\label{eq:ab10}
\end{equation}
and to the first order in $\xi_{k}$, 
\begin{equation}
K_{l}B_{2}=\mu K_{a}T_{ka}.
\label{eq:ab11}
\end{equation}
From the first equation of Eq. (\ref{eq:ab8}), we obtain
\begin{equation}
B_{1}=\bar{G}_{l}\xi_{k}=-f|_{z=0}\bar{G}_{l}\zeta_{k},
\label{eq:ab12}
\end{equation}
where we have used the relation Eq. (\ref{eq:se18}) in the dimensional form. Here $f(z)$ has the following form by substitution of $y_{*}=1-z$ into Eq. (\ref{eq:se17}):
\begin{eqnarray}
f(z)
&=&\frac{1}{6-i\alpha}(-6+i\alpha z+6z^{2}-i\alpha z^{3}) \nonumber \\
& &-\frac{\mu {\rm Re}\alpha}{210(6-i\alpha)^{2}}
\left\{144+(-174+5i\alpha)z-144z^{2} \right.
\nonumber \\
& &\left.+(210-11i\alpha)z^{3}+(-42+7i\alpha)z^{5}+(6-i\alpha)z^{7}\right\}.
\label{eq:ab13}
\end{eqnarray}
From the second equation of Eq. (\ref{eq:ab8}),
\begin{equation}
T_{ka}=\bar{G}_{a}\xi_{k}.
\label{eq:ab14}
\end{equation}
Eliminating $T_{ka}$ from Eqs. (\ref{eq:ab11}) and (\ref{eq:ab14}) and using Eq. (\ref{eq:ab10}), we obtain
\begin{equation}
B_{2}=\mu B_{1}.
\label{eq:ab15}
\end{equation}
By substitutions of Eqs. (\ref{eq:ab12}) and (\ref{eq:ab15}) into Eq. (\ref{eq:pt7}), we finally obtain
\begin{eqnarray}
g_{l}(z)&=&\left[-f|_{z=0}\left(\phi_{1}(z)+\mu\phi_{2}(z)\right)
+i\mu {\rm Pe} \int_{0}^{z}\left\{\phi_{2}(z)\phi_{1}(z')-\phi_{1}(z)\phi_{2}(z')\right\}f(z')dz'\right]\bar{G}_{l}\zeta_{k} \nonumber \\
&\equiv& H_{l}(z)\bar{G}_{l}\zeta_{k}.
\label{eq:ab16}
\end{eqnarray}

\section{Application}
In this section, we apply the solutions obtained in Sec. III and IV to the general formulae in Sec. II. If we assume $\bar{T}_{s}=T_{m}$ in the solid, Eq. (\ref{eq:b7}) becomes
\begin{equation}
L\bar{V}=K_{l}\bar{G}_{l}.
\label{eq:gd1}
\end{equation}
We solve Eq. (\ref{eq:g12}) in the quasi-stationary approximation $\sigma/(\kappa_{s}k^{2})\ll 1$ and $kl_{s} \gg 1$, where $l_{s}=\kappa_{s}/\bar{V}$ being the thermal diffusion length of the solid \cite{25}, and in the condition that the disturbance must vanish far from the solid-liquid interface, then the propagator in the solid is  
\begin{equation}
Q_{s}=\frac{\frac{dg_{s}}{dy}\Big|_{y=0}}{g_{s}|_{y=0}}=k.
\label{eq:gd2}
\end{equation}
If we are interested in the long wavelength region such that $Q_{l}\kappa_{l}/\bar{V}$, $Q_{s}l_{s} \gg 1$, using Eq. (\ref{eq:gd1}), Eq. (\ref{eq:b9}) becomes
\begin{eqnarray}
\sigma&=&\bar{V}Q_{l}\left(1+\frac{G(k)}{\bar{G}_{l}}\right)
          +n\bar{V}Q_{s}\frac{G(k)}{\bar{G}_{l}} \nonumber \\
      &=&\frac{\bar{V}}{h_{0}}\frac{\frac{dH_{l}}{dz}|_{z=1}}{H_{l}|_{z=1}}
         \left(1+\frac{G(k)}{\bar{G}_{l}}\right)
          +n\bar{V}k\frac{G(k)}{\bar{G}_{l}},
\label{eq:gd3}
\end{eqnarray}
where
\begin{equation}
H_{l}|_{z=1}=-f|_{z=0}(\phi_{1}|_{z=1}+\mu\phi_{2}|_{z=1})
             +i\mu {\rm Pe}I|_{z=1},
\label{eq:gd4}
\end{equation}
and
\begin{equation}
\frac{dH_{l}}{dz}\Big|_{z=1}=-f|_{z=0}
           \left(\frac{d\phi_{1}}{dz}\Big|_{z=1}
         +\mu\frac{d\phi_{2}}{dz}\Big|_{z=1}\right)
         +i\mu {\rm Pe}J|_{z=1},
\label{eq:gd5}
\end{equation}
and where
\begin{equation}
I(z)=\int_{0}^{z}\left\{\phi_{2}(z)\phi_{1}(z')-\phi_{1}(z)\phi_{2}(z')\right\}
f(z')dz',
\label{eq:gd6}
\end{equation}
which describes the disturbance of the steady state temperature distribution by fluid flow normal to the interface, and 
\begin{equation}
J(z)=\int_{0}^{z}\left\{\frac{d\phi_{2}(z)}{dz}\phi_{1}(z')
  -\frac{d\phi_{1}(z)}{dz}\phi_{2}(z')\right\}f(z')dz'.
\label{eq:gd7}
\end{equation}

In the absence of flow, we put $\rm Pe=0$ in Eqs. (\ref{eq:pt4}) and (\ref{eq:pt5}). If we expand Eqs. (\ref{eq:pt4}) and (\ref{eq:pt5}) with respect to the powers of $\mu$ up to infinity using the recursion relation by setting $a_{1}=\mu^{2}/2$ and $a_{2}=0$ (see Appendix A), we obtain at $z=1$
\begin{equation}
\phi_{1}|_{z=1}=1+\frac{\mu^{2}}{2!}+\frac{\mu^{4}}{4!}+\ldots,
\label{eq:dMS1}
\end{equation}
\begin{equation}
\phi_{2}|_{z=1}=1+\frac{\mu^{2}}{3!}+\frac{\mu^{4}}{5!}+\ldots.
\label{eq:dMS2}
\end{equation}
In the same way, the derivative of Eqs. (\ref{eq:pt4}) and (\ref{eq:pt5}) at $z=1$ are
\begin{equation}
\frac{d\phi_{1}}{dz}\Big|_{z=1}=\mu\left(\mu+\frac{\mu^{3}}{3!}
                                 +\frac{\mu^{5}}{5!}+\ldots\right),
\label{eq:dMS3}
\end{equation}
\begin{equation}
\frac{d\phi_{2}}{dz}\Big|_{z=1}=1+\frac{\mu^{2}}{2!}
                                 +\frac{\mu^{4}}{4!}+\ldots.
\label{eq:dMS4}
\end{equation}
Then the propagator in the liquid becomes
\begin{equation}
Q_{l}=\frac{1}{h_{0}}\frac{\frac{dH_{l}}{dz}|_{z=1}}{H_{l}|_{z=1}}=\frac{1}{h_{0}}\frac{\frac{d\phi_{1}}{dz}\Big|_{z=1}
       +\mu \frac{d\phi_{2}}{dz}\Big|_{z=1}}
         {\phi_{1}|_{z=1}+\mu \phi_{2}|_{z=1}}
     =k.
\label{eq:dMS5}
\end{equation}
If we take $G(k)$ in Eq. (\ref{eq:gd3}) as the Gibbs-Thomson effect \cite{25}:
\begin{equation}
G(k)=-\frac{T_{m}\Gamma}{L}k^{2},
\label{eq:dMS6}
\end{equation}
where $\Gamma$ is the solid-liquid interface tension, Eq. (\ref{eq:gd3}) reduces to the dispersion relation in the Mullins-Sekerka theory \cite{25}: 
\begin{equation}
\sigma_{r}=\bar{V}k\left[1-d_{0}\frac{\kappa_{l}}{\bar{V}}(1+n)k^{2}\right], 
\hspace{0.5cm}
\sigma_{i}=0,
\label{eq:dMS7}
\end{equation}
where $n=K_{s}/K_{l}$ and $d_{0}=T_{m}\Gamma C_{p}/L^{2}$ is the capillary length, $C_{p}$ being the specific heat at constant pressure, and from Eq. (\ref{eq:gd1}),  
\begin{equation}
\bar{V}=\frac{C_{p}\kappa_{l}(T_{m}-T_{la})}{Lh_{0}}.
\label{eq:dMS8}
\end{equation}
It should be noted that the $k$ term in front of the brackets in Eq. (\ref{eq:dMS7}) comes from Eqs. (\ref{eq:gd2}) and (\ref{eq:dMS5}). 
The $\mu$ term in Eq. (\ref{eq:dMS5}) appears as the result of the assumption that the heat in the air is transported by thermal diffusion.

On the other hand, in the presence of flow, in the long wavelength region of about 1cm which is the typical wavelength of the wavy pattern observed on the surface of icicles, we can neglect the Gibbs-Thomson effect as discussed in the article \cite{20}. From Eq. (\ref{eq:b4}),
\begin{equation}
g_{l}|_{z=1}=\left(1+\frac{G(k)}{\bar{G}_{l}}\right)\bar{G}_{l}\zeta_{k},
\label{eq:dU1}
\end{equation}
and noting $g_{l}|_{z=1}=H_{l}|_{z=1}\bar{G}_{l}\zeta_{k}$ from Eq. (\ref{eq:ab16}), the following relation must be satisfied:
\begin{equation}
G(k)=(H_{l}|_{z=1}-1)\bar{G}_{l}.
\label{eq:dU2}
\end{equation}
Then Eq. (\ref{eq:gd3}) can be written as
\begin{equation}
\sigma=\frac{\bar{V}}{h_{0}}\left\{{\frac{dH_{l}}{dz}}\Big|_{z=1}+n\mu
\left({H}_{l}|_{z=1}-1\right)\right\}.
\label{eq:dU3}
\end{equation}
We note that the second term on the right hand side of Eq. (\ref{eq:dU3}) represents the thermal diffusion of latent heat produced by a disturbed solid-liquid interface into the solid.
In the long wavelength region, we can make approximation of neglecting the $\mu^{2}$ term in Eqs. (\ref{eq:pt4}) and (\ref{eq:pt5}). This term is originated from the diffusion term $\partial^{2} T'_{l}/\partial x^{2}$ in Eq. (\ref{eq:g5}). In this case, the transport of heat in the liquid is dominated by shear flow. Noting that $\mu {\rm Pe}=u_{0}h_{0}^{2}k/\kappa_{l} \sim O(1)$ for the wavelength of about 1 cm observed on the surface of icicles when the typical values of $u_{0}\sim10^{-2}$ m/s, $h_{0}\sim10^{-4}$ m in the experiments \cite{21,22} and $\kappa_{l}=1.3\times 10^{-7}$ ${\rm m^{2}/s}$ of water are used \cite{20}, we expand Eqs. (\ref{eq:pt4}) and (\ref{eq:pt5}) with respect to the powers of $\mu \rm{Pe}$ up to the second order using the recursion relation by  setting $a_{1}=(\mu \rm Pe)^{1/2}/(2\sqrt{2})$ and $a_{2}=\mu \rm Pe/2-(\mu \rm Pe)^{1/2}/(2\sqrt{2})$ (see Appendix A) as follows: 
\begin{equation}
\phi_{1}(z)=1+\left(-\frac{1}{24}z^{4}+\frac{7}{360}z^{6}-\frac{1}{672}z^{8}\right)(\mu {\rm Pe})^2
+i\left(\frac{1}{2}z^{2}-\frac{1}{12}z^{4}\right)\mu \rm Pe,
\label{eq:dU4}
\end{equation}
\begin{equation}
\phi_{2}(z)=
z+\left(-\frac{1}{120}z^{5}+\frac{13}{2520}z^{7}
-\frac{1}{1440}z^{9}\right)(\mu {\rm Pe})^{2}
+i\left(\frac{1}{6}z^{3}-\frac{1}{20}z^{5}\right)\mu \rm Pe.
\label{eq:dU5}
\end{equation}
We evaluate each function and its derivative at $z=1$:
\begin{equation}
\phi_{1}|_{z=1}=1+i\frac{5}{12}\mu \rm Pe-\frac{239}{10080}(\mu \rm Pe)^2,
\label{eq:dU6}
\end{equation}
\begin{equation}
\phi_{2}|_{z=1}=1+i\frac{7}{60}\mu \rm Pe-\frac{13}{3360}(\mu \rm Pe)^2,
\label{eq:dU7}
\end{equation}
\begin{equation}
\frac{d\phi_{1}}{dz}\Big|_{z=1}=i\frac{2}{3}\mu \rm Pe-\frac{13}{210}(\mu \rm Pe)^2,
\label{eq:dU8}
\end{equation}
\begin{equation}
\frac{d\phi_{2}}{dz}\Big|_{z=1}=1+i\frac{1}{4}\mu \rm Pe-\frac{17}{1440}(\mu \rm Pe)^2,
\label{eq:dU9}
\end{equation}
and we evaluate Eq. (\ref{eq:ab13}) at $z=0$:
\begin{equation}
f|_{z=0}=\frac{-6}{6-i\alpha}
          -\frac{24\mu \rm Re\alpha}{35(6-i\alpha)^{2}}.
\label{eq:dU10}
\end{equation}

Substitutions of Eq. (\ref{eq:ab13}) and Eqs. (\ref{eq:dU4}) to (\ref{eq:dU9}) into Eqs. (\ref{eq:gd6}) and (\ref{eq:gd7}) and integration of them  give respectively at $z=1$,
 \begin{eqnarray}
i\mu {\rm Pe}I|_{z=1}&=&\frac{1}{36+\alpha^{2}}\left[\frac{9}{5}\alpha(\mu \rm Pe)+\left(\frac{239}{280}+\frac{13}{3360}\alpha^{2}\right)(\mu \rm Pe)^{2}
\right.\nonumber \\
& & \left. +i\left\{-\left(15+\frac{7}{60}\alpha^{2}\right)(\mu \rm Pe)
   +\frac{5}{42}\alpha (\mu \rm Pe)^{2}\right\}\right],
\label{eq:dU11}
\end{eqnarray}
\begin{eqnarray}
i\mu {\rm Pe}J|_{z=1}&=&\frac{1}{36+\alpha^{2}}\left[\frac{5}{2}\alpha(\mu \rm Pe)+\left(\frac{78}{35}+\frac{17}{1440}\alpha^{2}\right)(\mu \rm Pe)^{2}
\right.\nonumber \\
& & +\left.i\left\{-\left(24+\frac{1}{4}\alpha^{2}\right)(\mu \rm Pe)
   +\frac{101}{336}\alpha (\mu \rm Pe)^{2}\right\}\right],
\label{eq:dU12}
\end{eqnarray}
where we have carried out integration by neglect of the first order term in $\mu$ in Eq. (\ref{eq:ab13}) because this term is expected to give very small correction to Eqs. (\ref{eq:dU11}) and (\ref{eq:dU12}). 
By substitutions of Eqs. (\ref{eq:dU6}) to (\ref{eq:dU12}) into Eq. (\ref{eq:dU3}), the final forms of $\sigma_{r}$ and $v_{p}=-\sigma_{i}/k$ for the fluctuation of the solid-liquid interface in the long wavelength region are 
\begin{equation}
\sigma_{r}
=\frac{\bar{V}}{h_{0}}
\left[\frac{-\frac{3}{2}\alpha(\mu \rm Pe)
+\mu\left\{36-\frac{3}{2}\alpha(\mu \rm Pe)\right\}}{36+\alpha^{2}}
+n\mu
\frac{-\frac{7}{10}\alpha(\mu \rm Pe)+\mu\left\{36-\frac{7}{10}\alpha(\mu \rm Pe)\right\}-\alpha^{2}}{36+\alpha^{2}}\right],
\label{eq:dU13}
\end{equation}
\begin{equation}
v_{p}=-\frac{\bar{V}}{\mu}
\left[\frac{-\frac{1}{4}\alpha^{2}(\mu \rm Pe)
+\mu\left\{6\alpha+9(\mu \rm Pe)\right\}}{36+\alpha^{2}} 
+n\mu\frac{6\alpha-\frac{7}{60}\alpha^{2}(\mu \rm Pe)
+\mu\left\{6\alpha+\frac{21}{5}(\mu \rm Pe)\right\}}{36+\alpha^{2}}\right],
\label{eq:dU14}
\end{equation}
where we have neglected the first order term in $\mu$ in Eq. (\ref{eq:dU10}) by the same reason as mentioned above. Although we have expanded $\phi_{1}$ and $\phi_{2}$ up to the second order with respect to $\mu \rm Pe$ as in Eqs. (\ref{eq:dU4}) and (\ref{eq:dU5}), the values of coefficients of $(\mu \rm Pe)^{2}$ are very small compared to those of $\mu \rm Pe$. Indeed, we have confirmed that the form of $\sigma_{r}$ and $v_{p}$ including up to $(\mu \rm Pe)^{2}$ is almost the same as Eqs. (\ref{eq:dU13}) and (\ref{eq:dU14}) in the long wavelength region such that $k<10^{3}$ /m. Therefore, it is sufficient to approximate $\sigma_{r}$ and $v_{p}$ up to the first order in $\mu \rm Pe$. 


The rate of volume flow down the inclined plane in the experiment \cite{22} is  \begin{equation}
Q=u_{0}l\int_{0}^{h_{0}}\left(2\frac{y}{h_{0}}-\frac{y^{2}}{h_{0}^{2}}\right)dy=\frac{2}{3}u_{0}h_{0}l,
\label{eq:e1}
\end{equation}
where $l$ is the width of the gutter, and 
\begin{equation}
u_{0}=\frac{gh_{0}^{2}}{2\nu}\sin\theta
\label{eq:e2}
\end{equation}
is the surface velocity \cite{31}. If we eliminate $u_{0}$ from Eqs. (\ref{eq:e1}) and (\ref{eq:e2}), the mean thickness $h_{0}$ of the liquid can be expressed with respect to $Q$ and $\theta$:
\begin{equation}
h_{0}=\left(\frac{3\nu Q}{gl \sin\theta}\right)^{1/3}.
\label{eq:e3}
\end{equation}
Then, $\mu \rm Pe$ and $\alpha$ can be expressed in terms of $h_{0}$, respectively,
\begin{equation}
\mu {\rm Pe}=\frac{g \sin\theta}{2\kappa_{l}\nu}h_{0}^{4}k,
\label{eq:e4}
\end{equation}
\begin{equation}
\alpha=2 \cot\theta h_{0}k+a^{2}h_{0}k^{3},
\label{eq:e5}
\end{equation}
where we have defined the capillary constant associated with the surface tension $\gamma$ of the liquid-air \cite{31}: 
\begin{equation}
a=\sqrt{\frac{2\gamma}{g\rho_{l} \sin\theta}}.
\label{eq:e6}
\end{equation}
We note that this capillary constant depends on $\theta$, and that this typical value is about 3.9 mm for $\gamma=7.6\times 10^{-2}$ N/m and $\rho_{l}=1.0 \times 10^{3}$ $\rm{Kg/m^{3}}$ of water when $\theta=\pi/2$.

\begin{figure}
\begin{center}
\includegraphics[width=8cm,height=8cm,keepaspectratio,clip]{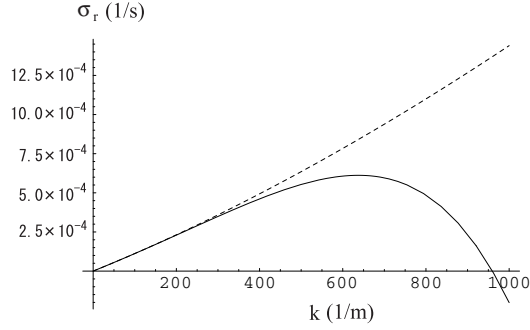}
\end{center}
\caption{The amplification rate $\sigma_{r}$ versus wave number $k$ for $T_{la}=-0.06^{\circ}$ C, $Q=160$ ml/hr and $\theta=\pi/2$. Solid line : with restoring forces. Dashed line : without restoring forces.}
\label{fig:amplification}
\end{figure}

\begin{figure}
\begin{center}
\includegraphics[width=8cm,height=8cm,keepaspectratio,clip]{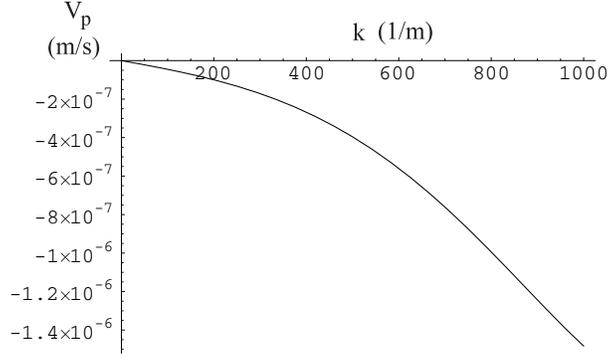}
\end{center}
\caption{The phase velocity $v_{p}=-\sigma_{i}/k$ versus wave number $k$ for $T_{la}=-0.06^{\circ}$ C, $Q=160$ ml/hr and $\theta=\pi/2$.}
\label{fig:phasevel}
\end{figure}

\begin{figure}
\begin{center}
\includegraphics[width=8cm,height=8cm,keepaspectratio,clip]{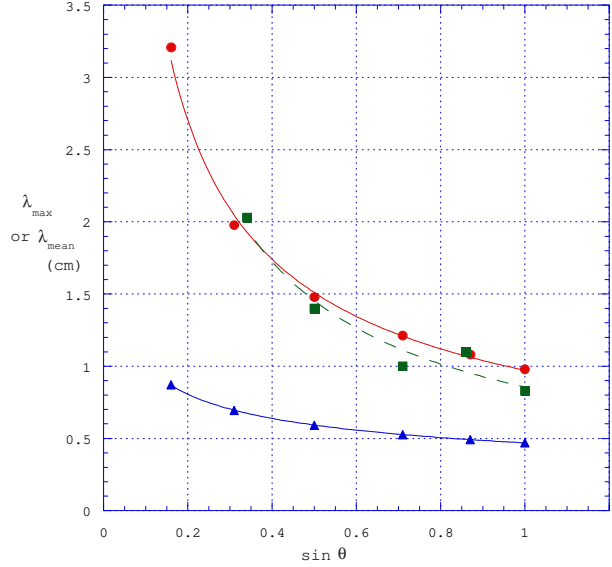}
\end{center}
\caption{The dependence of $\lambda_{max}$ or $\lambda_{mean}$ on the angle of inclined plane for $Q=160$ ml/hr. Closed circles : present result. Closed triangles : Ogawa-Furukawa's result \cite{20}. Closed squares : experimental result \cite{21}.}
\label{fig:wavelengths}
\end{figure}

From Fig. \ref{fig:amplification} to \ref{fig:wavelengths} we use the values of $T_{la}=-0.06^{\circ}$ C, $Q=160$ ml/hr and $l=0.03$ m in the experiments \cite{21,22}, and the physical properties of water, $L=3.3 \times 10^{8}$ ${\rm J/m^{3}}$, $C_{p}=4.2 \times 10^{6}$ ${\rm J/(K \hspace{1mm} m^{3})}$, $\kappa_{l}=1.3 \times 10^{-7}$ ${\rm m^{2}/s}$, $\nu=1.8 \times 10^{-6}$ ${\rm m^{2}/s}$, $\gamma=7.6\times 10^{-2}$ N/m and $n=K_{s}/K_{l}=3.92$, where $K_{s}$ is the thermal conductivity of ice. The reason for choosing the value of $Q=160$ ml/hr is that the clearest wavy pattern was observed at this value in the experiment \cite{22}. Since the crystal growth velocity $\bar{V}$ observed in the actual experiment is about $10^{-6}$ m/s \cite{21}, from Eq. (\ref{eq:dMS8}) we obtain the value of $T_{la}=-0.06^{\circ}$ C for water when $h_{0}=10^{-4}$m. Although $T_{la}$ is to be determined by the condition of the surrounding air, we use this value for $T_{la}$ to determine the value of $\bar{V}$ from Eq. (\ref{eq:dMS8}) when varing $\theta$. 

The solid line in Fig. \ref{fig:amplification} shows the amplification rate Eq. (\ref{eq:dU13}) versus wave number $k$ for $\theta=\pi/2$. This shows that $\sigma_{r}$ takes a maximum value $\sigma_{r max}$ at a wave number. The characteristic time for most unstable mode is $1/\sigma_{r max}$ and is about 30 minutes in this case. Indeed, it is reported in the experiment that a periodic structure as the original form of wavy pattern is observed in about 30 minutes \cite{22}. On the other hand, the dashed line in Fig. \ref{fig:amplification} shows $\sigma_{r}$ when we neglect the restoring forces due to gravity and surface tension, i.e., when we put $\alpha=0$ in Eq. (\ref{eq:dU13}). Then $\sigma_{r}$ is always positive in the range of our interests.

Fig. \ref{fig:phasevel} shows the phase velocity Eq. (\ref{eq:dU14}) versus wave number $k$ for $\theta=\pi/2$. This shows that the fluctuation of the solid-liquid interface for the maximum point of $\sigma_{r}$ in Fig. \ref{fig:amplification} moves upward along the icicle with the magnitude of about 0.6 $\bar{V}$. Indeed, there is evidence to support our predictions that many tiny air bubbles dissolved in the thin flowing liquid are trapped in just upstream region of any protrudent part on a growing icicle and its region migrates in the upward direction during growth (see Fig. 9B in Ref. \cite{21}). This suggests that the velocity of ice growth is faster in the upstream region of each protrudent part.
On the other hand, in the O-F model, it was predicted that the fluctuation moves downward along the icicle with the phase velocity, 
\begin{equation}
v_{p}=\bar{V}
\frac{\frac{5}{12}\mu \rm Pe}
{\left\{1-\frac{239}{10080}(\mu \rm Pe)^{2}\right\}^{2}
+\left\{\frac{5}{12}\mu \rm Pe\right\}^{2}}.
\label{eq:e7}
\end{equation}
If this prediction is correct, air bubbles would be trapped in the downstream region of each protrudent part and migrate in the downward direction during growth.

Fig. \ref{fig:wavelengths} shows the dependence of the wavelength $\lambda_{max}$ obtained theoretically or the mean wavelength $\lambda_{mean}$ obtained in the experiment on the angle $\theta$ of the inclined plane. The closed squares represent mean wavelength obtained by the experiment \cite{22}:
\begin{equation}
\lambda_{mean}\sim\frac{0.83}{(\sin\theta)^{0.6\sim0.9}} \hspace{0.3cm} {\rm (cm)}.
\label{eq:e8}
\end{equation} 
Using Eq. (\ref{eq:dU13}), we determine the wavelength $\lambda_{max}$ at the maximum point of $\sigma_{r}$, and its dependence on the angle is found to be
\begin{equation}
\lambda_{max}\sim\frac{0.98}{(\sin\theta)^{0.6\sim0.65}} \hspace{0.3cm} {\rm (cm)},
\label{eq:e9}
\end{equation}
which is shown in the closed circles. We note that this dependence of $\lambda_{max}$ on $\sin \theta$ comes from not only $\mu \rm Pe$ of Eq. (\ref{eq:e4}) but also $\cot\theta$, $h_{0}$ and $a$ in Eq. (\ref{eq:e5}). Our results are in good agreement with experiment.
On the other hand, the amplification rate obtained in the O-F model is 
\begin{equation}
\sigma_{r}=\bar{V}k
\frac{1-\frac{239}{10080}(\mu \rm Pe)^{2}}
{\left\{1-\frac{239}{10080}(\mu \rm Pe)^{2}\right\}^{2}
+\left\{\frac{5}{12}\mu \rm Pe\right\}^{2}}.
\label{eq:e10}
\end{equation}
The closed triangles are $\lambda_{max}$ at the maximum point of Eq. (\ref{eq:e10}). Then the result is 
\begin{equation}
\lambda_{max}\sim\frac{0.47}{(\sin\theta)^{1/3}}  \hspace{0.3cm} {\rm (cm)}.\label{eq:e11}
\end{equation}
In this case, we note that the dependence of $\lambda_{max}$ on $\sin \theta$ comes from only $\mu \rm Pe$ of Eq. (\ref{eq:e4}). 

\section{Discussion}
Some differences between our results and their results \cite{20} appear to arise from the following reasons. A main difference is originated from the order estimate of Eq. (\ref{eq:se12}) or Eq. (\ref{eq:e5}). If we use the values of $h_{0}$ at $Q=160$ ml/hr and $k=2\pi/\lambda_{max}$, where $\lambda_{max}$ is taken from Eq. (\ref{eq:e9}), the values of $\alpha$ and $\mu$ for $0.1<\sin \theta \leq 1$ take a range, $0.4<\alpha<0.8$ and $0.03<\mu<0.06$, respectively. Therefore, we have treated $\alpha$ as zeroth order in terms of $\mu$. On the other hand, it was regarded as first order in $\mu$ in the O-F model. When we determine the perturbed stream function, these differences cause different forms between Eqs. (\ref{eq:se17}) and (\ref{eq:se19}). As a result of that, different dependence of $\lambda_{\rm max}$ on $\sin\theta$  between Eqs. (\ref{eq:e9}) and (\ref{eq:e11}) has occured. The $\sin \theta$ term in Eq. (\ref{eq:dU13}) is included not only in $\mu \rm Pe$ but also $\alpha$. On the other hand, in the O-F model, the $\sin \theta$ term in Eq. (\ref{eq:e10}) appears in only $\mu \rm Pe$. If we use Eq. (\ref{eq:se17}), in which the effect of restoring forces due to gravity and surface tension on the liquid-air surface is included, $\sigma_{r}$ takes a maximum value at a wave number. On the other hand, if we use Eq. (\ref{eq:se19}), in which the effect of restoring forces is not included, $\sigma_{r}$ is always positive in the long wavelength region. In spite of absence of the $\alpha$ term in the O-F model, the similar curve as solid line in Fig. \ref{fig:amplification} was obtained (see Fig. 4 in Ref. \cite{20}). The existence of maximum of their $\sigma_{r}$ is the result of expansion of the temperature fluctuation in the liquid up to $(\mu \rm Pe)^{2}$, for example, which is reflected in the numerator in Eq. (\ref{eq:e10}). However, we have confirmed that it is sufficient to approximate $\sigma_{r}$ up to the first order in $\mu \rm Pe$. The existence of maximum of our $\sigma_{r}$ comes from the effect of $\alpha$ and $\mu \rm Pe$.

These different results and the different prediction of the direction of the phase velocity between Eqs. (\ref{eq:dU14}) and (\ref{eq:e7}) may be due to the difference of the boundary condition for the temperature at the solid-liquid interface and that of the liquid-air surface between ours and theirs. Instead of Eqs. (\ref{eq:b1}) and (\ref{eq:ab6}), the following boundary conditions were used in the O-F model, respectively,
\begin{equation}
(\bar{T}_{l}+T'_{l})|_{y=\zeta}=(\bar{T}_{s}+T'_{s})|_{y=\zeta}=T_{m},
\label{eq:dis1}
\end{equation}
\begin{equation}
(\bar{T}_{l}+T'_{l})|_{y=\xi}=(\bar{T}_{a}+T'_{a})|_{y=\xi}.
\label{eq:dis2}
\end{equation}
In order to determine the two constants $B_{1}$ and $B_{2}$ in Eq. (\ref{eq:pt7}) independently, and in the absence of flow, to recover the usual Mullins-Sekerka theory from the general dispersion relation Eq. (\ref{eq:gd3}), we have used the boundary condition Eq. (\ref{eq:ab6}) instead of Eq. (\ref{eq:dis2}). For long wavelength fluctuation of about 1 cm of the solid-liquid interface, since we can neglect the change of the melting temperature due to the Gibbs-Thomson effect, adopting Eq. (\ref{eq:dis1}) appears to be appropriate. In the presence of flow, however, Eq. (\ref{eq:dU2}) suggests that there exists a shift of the melting temperature depending on the wave number. Therefore we have used the boundary condition Eq. (\ref{eq:b1}) instead of Eq. (\ref{eq:dis1}).

$\mu$ term in the numerator in Eq. (\ref{eq:dU13}) is the cause of instability. This term is originated from spatial derivative of the perturbed air temperature distribution at the deformed liquid-air surface as indicated on the right hand side of Eq. (\ref{eq:ab11}). From Eq. (\ref{eq:se12}) or Eq. (\ref{eq:e5}), since the value of $\alpha$ is very small in the long wavelength region, the effect of restoring forces due to gravity and surface tension on the liquid-air surface is small. Therefore, in the low wave number region instability effect with positive terms in Eq. (\ref{eq:dU13}) dominates stability effect with negative terms. On the other hand, as increasing the wave number, since the value of $\alpha$ increases, the effect of restoring forces is large. Then $\alpha(\mu \rm Pe)$ and $\alpha^{2}$ terms with negative sign in the numerator of Eq. (\ref{eq:dU13}) dominate the instability terms. As a result of that, we obtain the solid curve in Fig. \ref{fig:amplification}. In order to interpret correctly the physical mechanism of instability and stability of the solid-liquid interface and why the solid-liquid interface moves in the upstream direction, it is necessary to understand the relative phase of modes at each interface using the relation Eq. (\ref{eq:se18}) and a shift of melting temperature due to flow and restoring forces as suggested in Eq. (\ref{eq:dU2}). This will be shortly clarified in another paper. 

In Sec. III and IV, we have made some assumptions. Here we justify them. We have assumed the time independence of the purterbed flow, therefore we have neglected the $\sigma_{*}/\mu$ term in Eqs. (\ref{eq:f23}), (\ref{eq:fb6}), (\ref{eq:fb8}) and (\ref{eq:fb10}). This assumption is valid because we see from Fig. \ref{fig:amplification} and Fig. \ref{fig:phasevel} that $\sigma_{*}\sim10^{-6}$ for $\bar{V} \sim 10^{-6}$ m/s. Therefore the condition $\sigma_{*}/\mu \ll 1$ is satisfied. The same can be said for the fluctuation of the temperature. For a deformation of wave number $k$, the characteristic delay time of the fluctuation of the temperature is of order $\Delta t_{thermal} \sim (\kappa_{l} k^2)^{-1}$. $\Delta t_{thermal}$ is much smaller than the characteristic time of evolution of the mode, $\sigma_{r}^{-1}$. Therefore the condition $\sigma_{r}/(\kappa_{l}k^{2}) \ll 1$ is satisfied. These mean that the perturbed flow field and the perturbed temperature field respond relatively rapidly to the slow development of the solid-liquid interface. 

\section{Conclusion}
The restoring forces due to gravity and surface tension determine the shape of free surface and do not directly act on the solid-liquid interface. However, the effect of restoring forces has played an important role on stabilization of the solid-liquid interface. Although the Gibbs-Thomson effect acts effectivley on the micrometer scale, we have found that the effect of restoring forces is more effective for long wavelength fluctuation of the order of mm, which is of the order of the capillary constant associated with the surface tension of the liquid-air. Therefore the wavy pattern observed on the surface of icicles and inclined plane occures on longer length scales compared to the wavelength predicted by the usual Mullins-Sekerka theory.   

Since our calculations are based on the linear stability analysis, our formulations and the Ogawa and Furukawa's formulations do not have direct correspondences. The relation between our present formulations and previous ones is under investigation by Ogawa and Furukawa. 

\begin{acknowledgements}
The author would like to thank N. Ogawa and M. Sato for very useful comments and remarks about the paper. The author would also like to thank Y. Furukawa and E. Yokoyama for helpful discussions. 
\end{acknowledgements}

\appendix

\section{ }

We separate Eq. (\ref{eq:pt4}) into the real part and imaginary part:
\begin{eqnarray}
\phi_{1}(z)&=&\exp\left(-\frac{(\mu \rm Pe)^{1/2}}{2\sqrt{2}}z^{2}\right)
\left[\cos\left(\frac{(\mu \rm Pe)^{1/2}}{2\sqrt{2}}z^{2}\right)(1+\sum_{j=0}^{\infty}a_{2j+1}z^{2j+2}) \right. \nonumber \\
       & &-\sin\left(\frac{(\mu \rm Pe)^{1/2}}{2\sqrt{2}}z^{2}\right)(\sum_{j=0}^{\infty}a_{2j+2}z^{2j+2}) \nonumber \\
& & +i\left\{\cos\left(\frac{(\mu \rm Pe)^{1/2}}{2\sqrt{2}}z^{2}\right)(\sum_{j=0}^{\infty}a_{2j+2}z^{2j+2}) \right.\nonumber \\
& & \left.\left.+\sin\left(\frac{(\mu \rm Pe)^{1/2}}{2\sqrt{2}}z^{2}\right)(1+\sum_{j=0}^{\infty}a_{2j+1}z^{2j+2})\right\}\right].
\label{eq:appa1}
\end{eqnarray}
Using the first two coefficients,
\begin{equation}
a_{1}=\frac{1}{2}\mu^{2}+\frac{1}{2\sqrt{2}}(\mu \rm Pe)^{1/2}, 
\label{eq:appa2}
\end{equation}
\begin{equation}
a_{2}=\frac{1}{2}\mu \rm Pe-\frac{1}{2\sqrt{2}}(\mu \rm Pe)^{1/2},
\label{eq:appa3}
\end{equation}
the other coefficients $a_{j}$ for odd numbers are obtained from the following recursion relation, 
\begin{equation}
a_{2j+1}=\frac{1}{(j+1)(2j+1)}\left\{a_{1}a_{2j-1}-a_{2}a_{2j}+\frac{2j}{\sqrt{2}}(\mu {\rm Pe})^{1/2}(a_{2j-1}+a_{2j})\right\} \hspace{5mm} (j=1,2,3,\ldots),
\label{eq:appa4}
\end{equation}
and for even numbers,
\begin{equation}
a_{2j+2}=\frac{1}{(j+1)(2j+1)}\left\{a_{1}a_{2j}+a_{2}a_{2j-1}-\frac{2j}{\sqrt{2}}(\mu {\rm Pe})^{1/2}(a_{2j-1}-a_{2j})\right\} \hspace{5mm} (j=1,2,3,\ldots).
\label{eq:appa5}
\end{equation}


Next, we separate the derivative of Eq. (\ref{eq:pt4}) into the real part and imaginary part:
\begin{eqnarray}
\frac{d\phi_{1}(z)}{dz}&=&\exp\left(-\frac{(\mu \rm Pe)^{1/2}}{2\sqrt{2}}z^{2}\right)
\left[\cos\left(\frac{(\mu \rm Pe)^{1/2}}{2\sqrt{2}}z^{2}\right)(\sum_{j=0}^{\infty}b_{2j+1}z^{2j+1}) \right.\nonumber \\
       & &-\sin\left(\frac{(\mu \rm Pe)^{1/2}}{2\sqrt{2}}z^{2}\right)(\sum_{j=0}^{\infty}b_{2j+2}z^{2j+1}) \nonumber \\
& & +i\left\{\cos\left(\frac{(\mu \rm Pe)^{1/2}}{2\sqrt{2}}z^{2}\right)(\sum_{j=0}^{\infty}b_{2j+2}z^{2j+1}) \right.\nonumber \\
& &+\left.\left.\sin\left(\frac{(\mu \rm Pe)^{1/2}}{2\sqrt{2}}z^{2}\right)(\sum_{j=0}^{\infty}b_{2j+1}z^{2j+1})\right\}\right].
\label{eq:appb1}
\end{eqnarray}
Using
\begin{equation}
b_{1}=2a_{1}-\frac{1}{\sqrt{2}}(\mu \rm Pe)^{1/2}, 
\label{eq:appb2}
\end{equation}
\begin{equation}
b_{2}=2a_{2}+\frac{1}{\sqrt{2}}(\mu \rm Pe)^{1/2},
\label{eq:appb3}
\end{equation}
the other coefficients $b_{j}$ for odd numbers are obtained from 
\begin{equation}
b_{2j+1}=2(j+1)a_{2j+1}-\frac{1}{\sqrt{2}}(\mu {\rm Pe})^{1/2}(a_{2j-1}+a_{2j}) \hspace{5mm} (j=1,2,3,\ldots),
\label{eq:appb4}
\end{equation}
and for even numbers,
\begin{equation}
b_{2j+2}=2(j+1)a_{2j+2}-\frac{1}{\sqrt{2}}(\mu {\rm Pe})^{1/2}(a_{2j-1}-a_{2j}) \hspace{5mm} (j=1,2,3,\ldots).
\label{eq:appb5}
\end{equation}


Eq. (\ref{eq:pt5}) is separated into the real part and imaginary part in the same way:
\begin{eqnarray}
\phi_{2}(z)&=&\exp\left(-\frac{(\mu \rm Pe)^{1/2}}{2\sqrt{2}}z^{2}\right)
\left[\cos\left(\frac{(\mu \rm Pe)^{1/2}}{2\sqrt{2}}z^{2}\right)(z+\sum_{j=0}^{\infty}c_{2j+1}z^{2j+3}) \right.\nonumber \\
       & &-\sin\left(\frac{(\mu \rm Pe)^{1/2}}{2\sqrt{2}}z^{2}\right)(\sum_{j=0}^{\infty}c_{2j+2}z^{2j+3}) \nonumber \\
& & +i\left\{\cos\left(\frac{(\mu \rm Pe)^{1/2}}{2\sqrt{2}}z^{2}\right)(\sum_{j=0}^{\infty}c_{2j+2}z^{2j+3}) \right.\nonumber \\
& &+\left.\left.\sin\left(\frac{(\mu \rm Pe)^{1/2}}{2\sqrt{2}}z^{2}\right)(z+\sum_{j=0}^{\infty}c_{2j+1}z^{2j+3})\right\}\right].
\label{eq:appc1}
\end{eqnarray}
Using
\begin{equation}
c_{1}=\frac{1}{3}a_{1}+\frac{1}{3\sqrt{2}}(\mu \rm Pe)^{1/2}, 
\label{eq:appc2}
\end{equation}
\begin{equation}
c_{2}=\frac{1}{3}a_{2}-\frac{1}{3\sqrt{2}}(\mu \rm Pe)^{1/2}, 
\label{eq:appc3}
\end{equation}
the other coefficients $c_{j}$ for odd numbers are obtained from
\begin{equation}
c_{2j+1}=\frac{1}{(j+1)(2j+3)}\left\{a_{1}c_{2j-1}-a_{2}c_{2j}+\frac{2j+1}{\sqrt{2}}(\mu {\rm Pe})^{1/2}(c_{2j-1}+c_{2j})\right\} \hspace{5mm} (j=1,2,3,\ldots),\label{eq:appc4}
\end{equation} 
and for even numbers,
\begin{equation}
c_{2j+2}=\frac{1}{(j+1)(2j+3)}\left\{a_{1}c_{2j}+a_{2}c_{2j-1}-\frac{2j+1}{\sqrt{2}}(\mu {\rm Pe})^{1/2}(c_{2j-1}-c_{2j})\right\} \hspace{5mm} (j=1,2,3,\ldots).\label{eq:appc5}
\end{equation}


Finally, derivative of $\phi_{2}$ is separated into the real part and imaginary part:
\begin{eqnarray}
\frac{d\phi_{2}(z)}{dz}&=&\exp\left(-\frac{(\mu \rm Pe)^{1/2}}{2\sqrt{2}}z^{2}\right)
\left[\cos\left(\frac{(\mu \rm Pe)^{1/2}}{2\sqrt{2}}z^{2}\right)(1+\sum_{j=0}^{\infty}d_{2j+1}z^{2j+2}) \right.\nonumber \\
       & &-\sin\left(\frac{(\mu \rm Pe)^{1/2}}{2\sqrt{2}}z^{2}\right)(\sum_{j=0}^{\infty}d_{2j+2}z^{2j+2}) \nonumber \\
& & +i\left\{\cos\left(\frac{(\mu \rm Pe)^{1/2}}{2\sqrt{2}}z^{2}\right)(\sum_{j=0}^{\infty}d_{2j+2}z^{2j+2}) \right.\nonumber \\
& &+\left.\left. \sin\left(\frac{(\mu \rm Pe)^{1/2}}{2\sqrt{2}}z^{2}\right)(1+\sum_{j=0}^{\infty}d_{2j+1}z^{2j+2})\right\}\right],
\label{eq:appd1}
\end{eqnarray}
where $d_{1}=a_{1}$, $d_{2}=a_{2}$,
and other coefficients $d_{j}$ for odd numbers are obtained from
\begin{equation}
d_{2j+1}=(2j+3)c_{2j+1}-\frac{1}{\sqrt{2}}(\mu {\rm Pe})^{1/2}(c_{2j-1}+c_{2j}) \hspace{5mm} (j=1,2,3,\ldots),
\label{eq:appd2}
\end{equation}
and for even numbers,
\begin{equation}
d_{2j+2}=(2j+3)c_{2j+2}-\frac{1}{\sqrt{2}}(\mu {\rm Pe})^{1/2}(c_{2j-1}-c_{2j}) \hspace{5mm} (j=1,2,3,\ldots).
\label{eq:appd3}
\end{equation}

It should be noted that all coefficients are obtained from only $a_{1}$ and $a_{2}$. Eqs. (\ref{eq:dU6}) to (\ref{eq:dU9}) are valid only in the long wavelength region because we neglect the $\mu^{2}$ term in Eq. (\ref{eq:appa2}). This means that heat transport is dominated by shear flow. On the other hand, in the absence of flow, we put $\rm Pe=0$ in  Eq. (\ref{eq:appa2}). Then Eqs. (\ref{eq:dMS1}) to (\ref{eq:dMS4}) are obtained. In this case, heat transport is dominated by thermal diffusion.

\end{document}